\def\bm#1{\mbox{\boldmath $#1$}}

\documentstyle [12pt]{article}

\def\nsection#1{\section{#1}\setcounter{equation}{0}}

\begin{document}

\hoffset = -1truecm
\voffset = -2truecm

\thispagestyle{empty}
\begin{flushright}
{\large \bf DFTUZ/94/26}\\
{\large \bf hep-th/9412137}
\end{flushright}
\vskip 3truecm

\begin{center}

{\large \bf   DIRAC VERSUS REDUCED PHASE SPACE QUANTIZATION\footnote{Talk
presented at the Geometry of Constrained Dynamical Systems Conference,
Cambridge, 15-18 June 1994}}\\
\vskip0.8cm
{ \bf
Mikhail S. Plyushchay${}^a$\footnote{On leave from the
Institute for High Energy Physics,
Protvino, Moscow Region, Russia; E-mail: mikhail@cc.unizar.es}
and Alexander V. Razumov${}^b$\footnote{E-mail: razumov@mx.ihep.su}
}\\[0.3cm]
{\it ${}^a$Departamento de F\'{\i}sica Te\'orica, Facultad de Ciencias}\\
{\it Universidad de Zaragoza, Zaragoza 50009, Spain}\\
{\it ${}^b$ Institute for High Energy Physics, Protvino}\\
{\it Moscow Region, 142284 Russia}\\

\vskip2cm

                            {\bf Abstract}
\end{center}
The relationship between the Dirac and reduced phase space quantizations is
investigated for spin models belonging to the class of Hamiltonian systems
having no gauge conditions.  It is traced out that the two quantization
methods may give similar, or essentially different physical results, and,
moreover, it is shown that there is a class of constrained systems, which can
be quantized only by the Dirac method.  A possible interpretation of the
gauge degrees of freedom is given.

\vfill
\newpage

\nsection{Introduction}

There are two main methods to quantize the Hamiltonian systems with first
class constraints: the Dirac quantization \cite{Dir} and the reduced phase
space quantization \cite{Fad69}, whereas two other methods, the path
integral method \cite{FPo67,Fad69} and the BRST quantization \cite{brst}
being the most popular method for the
covariant quantization of gauge-invariant systems, are based on and
proceed from them \cite{Fad69,sun}.  The basic idea of the Dirac
method consists in imposing quantum mechanically the first class
constraints as operator conditions on the states for singling out the
physical ones \cite{Dir}. The reduced phase
space quantization first identifies
the physical degrees of freedom at the classical level by the
factorization of the constraint surface with respect to the action of the
gauge group, generated by the constraints. Then the resulting Hamiltonian
system is quantized as a usual unconstrained system \cite{Fad69}.
Naturally, the problem of the relationship of these two methods arises. It
was discussed in different contexts in literature \cite{kuc}, and
there is an opinion that the differences
between the two quantization methods can be traced out to a choice of
factor ordering in the construction of various physical operators.

We investigate the relationship of the two methods of
quantization for the special class of Hamiltonian systems with first class
constraints corresponding to different physical models of spinning
particles.  The specific general property of the examples of
constrained systems considered here is the following:
their constraints generate SO(2)
transformations and, hence, corresponding gauge orbits topologically are
one-spheres $S^{1}$. This fact implies that these systems {\it do not
admit gauge conditions}, and, therefore, for the construction
of their reduced phase spaces
we shall use a general geometrical approach to
the Dirac--Bergmann theory of the constrained systems \cite{AbM78,PR}.

\nsection{Plane Spin Model}

The first model we are going to consider is the plane spin model,
which is a subsystem of the (3+1)--dimensional models of massless
particles with arbitrary helicity \cite{ply2},
and of the (2+1)--dimensional relativistic
models of fractional spin particles \cite{ply3}.
The initial phase space of the model is a cotangent bundle
$T^* S^1$ of the one--dimensional sphere $S^1$, that is a cylinder
$S^{1}\times {\bf R}$. It can be described {\it locally}
by an angular variable
$0\le \varphi < 2\pi$ and the conjugate momentum $S\in {\bf R}$.
The symplectic
two--form $\omega$ in terms of the local variables $\varphi$, $S$ has the
form
$
\omega = dS \wedge d \varphi,
$
and, thus, we have locally
$
\{\varphi, S\} = 1.
$
Actually, any $2\pi$--periodical function of the variable $\varphi$ that
is considered as a variable, taking values in ${\bf R}$, can be considered
as a function on the phase space, i.e., as an observable, and any
observable is connected with the corresponding $2\pi$--periodical
function. Therefore, we can introduce the functions
$q_1 = \cos \varphi,$ $q_2 = \sin \varphi$,
$q_1^2 + q_2^2 = 1$,
as the dependent functions on the phase space of the system.
For these functions we have
$\{q_1, q_2\}= 0,$
$\{q_{1}, S\} = -q_{2},$
$\{q_{2}, S\} = q_{1}.$
Any function on the phase space can be considered as a function
of dependent coordinates $q_1$, $q_2$ and $S$, which will be
taken below as the quantities, forming a restricted set of observables
whose quantum analogs have the commutators which are in the direct
correspondence with their Poisson brackets.

We come to the plane spin model by introducing the `spin' constraint
\begin{equation}
\psi = S - \theta = 0, \label{2.9}
\end{equation}
where $\theta$ is an arbitrary real constant.
Let us consider the Dirac quantization of the system.
To this end we take
as the Hilbert space the space of complex $2\pi$--periodical functions of
the variable $\varphi$ with the scalar pro\-duct
$(\Phi_1,\Phi_2) = \frac{1}{2\pi} \int_0^{2\pi}
\overline{\Phi_1(\varphi)}\Phi_2 (\varphi)\, d\varphi.$
The operators $\hat q_1$ and $\hat q_2$, corresponding to the functions
$q_1$ and $q_2$, are the operators of multiplication by the functions
$\cos \varphi$ and $\sin\varphi$, respectively, whereas
the operator $\hat S$ is defined by
$
\hat S \Phi = (-id/d \varphi + c)\Phi,
$
where $c$ is an arbitrary real constant. The operators
$\hat q_1$, $\hat q_2$ and $\hat S$ are Hermitian operators with respect
to the introduced scalar product, and they satisfy the
relation $[\hat{A},\hat{B}]=i\{A,B\}$, $A,B=q_1 ,q_2 ,S$.
The quantum analog of the constraint (\ref{2.9}) gives the equation for
the physical state wave functions:
$(\hat S~-~\theta)\Phi_{phys} = 0.$
Decomposing the function $\Phi_{phys}(\varphi)$ over the orthonormal basis,
formed by the functions $e^{ik\varphi}$, $k\in {\bf Z}$,
we find this equation has a nontrivial solution only when
$c = \theta + n,$
where $n$ is some fixed integer, $n \in {\bf Z}.$ In this case the
corresponding physical normalized wave function is
$\Phi_{phys}(\varphi) = e^{in\varphi}.$
The only physical operator \cite{sun}, i.e., an operator commuting
with the quantum constraint $\hat \psi$ here is $\hat S$,
which is reduced to the
constant $\theta$ on the physical subspace.

Now we come back to the classical theory in order to construct the reduced
phase space of the model. Let us show that for the surface, defined by
Eq.~(\ref{2.9}), there is no `good' gauge condition, but, nevertheless, the
reduced phase space of the system can be constructed. Indeed, it is clear
that the one--parameter group of transformations, generated by the
constraint $\psi$, consists of the rotations of the phase space. This group
acts transitively on the constraint surface, and we have only one gauge
orbit, which is the constraint surface itself. The gauge conditions must
single out one point of an orbit. In our case we have to define only one
gauge condition, let us denote it by $\chi$. The function $\chi$ must be
such that the pair of equations
$\psi = 0,$
$\chi = 0$
would determine a set, consisting of only one point, and in this point we
should have
$\{\psi, \chi\} \ne 0.$
Recall that any function on the phase space of the system
under consideration can be considered as a function of the variables
$\varphi$ and $S$, which is $2\pi$--periodical with respect to $\varphi$.
Thus, we require the $2\pi$--periodical function
$\chi(\varphi, S)$
turn into zero at only one point
$\varphi = \varphi_0$ from the interval $0 \le \varphi < 2\pi$
when $S=\theta$. Moreover,
we should have
$
\{\psi, \chi\}(\varphi_0, \theta) = - \left. \partial \chi(\varphi,
\theta)/\partial \varphi\right|_{\varphi = \varphi_0} \ne 0.
$
It is clear that such a function does not exist.  Nevertheless, here we have
the reduced phase space that consists of only one point. Therefore,
the reduced space quantization is trivial: physical operator $\hat S$
takes here constant value $\theta$ in correspondence with the results
obtained by the Dirac quantization method.  When the described plane spin
model is a subsystem of some other system, the reduction means simply that
the cylinder $T^{*}S^{1}$ is factorized into a point, where $S = \theta$,
and that wave functions do not depend on the variable $\varphi$.

Let us point out one interesting analogy in interpretation of the
situation with nonexistence of a global gauge condition. Here the
condition of $2\pi$--periodicity can be considered as a `boundary'
condition. If for a moment we forget about it, we can take as a gauge
function any monotonic function $\chi(\varphi, S)$, $\chi \in {\bf R}$,
such that $\chi(\varphi_{0}, \theta)=0$ at some point $\varphi =
\varphi_{0}$, and, in particular, we can choose the function
$\chi(\varphi, S) = \varphi$. The `boundary' condition excludes all such
global gauge conditions. In this sense the situation is similar to the
situation in the non--Abelian gauge theories where without taking into
account the boundary conditions for the fields it is also possible to find
global gauge conditions, whereas the account of those leads, in the end, to
the nonexistence of global gauge conditions \cite{sin}.

\nsection{Rotator Spin Model}

Let us consider now the rotator spin model \cite{ply4}. The
initial phase space of the system is described by a spin three--vector
${\bm S}$ and a unit vector ${\bm q}$,
$
{\bm q}^2=1,
$
being orthogonal one to the other,
$
{\bm q}{\bm S} = 0.
$
The variables $q_i$ and $S_i$, $i=1,2,3$, can be considered as dependent
coordinates in the phase space of the system. The Poisson brackets for
these coordinates are
$\{q_i, q_j\} = 0,$ $\{S_i, S_j\} = \epsilon_{ijk}S_k,$
$\{S_i,q_j\} = \epsilon_{ijk} q_k.$
Using these Poisson brackets,
we find the following expression for the symplectic two--form:
$
\omega = d p_i \wedge d q_i = d(\epsilon_{ijk} S_j q_k) \wedge d q_i.
$
Introducing the spherical angles $\varphi$, $\vartheta$ ($0 \le \varphi <
2\pi,\, 0 \le \vartheta \le \pi)$ and the corresponding momenta
$p_\varphi, p_\vartheta \in {\bf R}$, we can write the
parameterization for the vector $\bm q$,
$\bm q=(\cos \varphi \sin \vartheta,
 \sin \varphi \sin \vartheta,\cos \vartheta)$
and corresponding parameterization for the vector $\bm S=\bm S(\vartheta,
\varphi, p_\varphi, p_\vartheta)$, whose explicit form we do not
write down here (see Ref.~\cite{PR}).
Then for the symplectic two--form we get the expression
$
\omega = d p_\vartheta \wedge d \vartheta + d p_\varphi \wedge d\varphi.
$
{}From this relation we conclude that the initial phase space of the system
is symplectomorphic to the cotangent bundle $T^*S^2$ of the
two--dimensional sphere $S^2$, furnished with the canonical symplectic
structure.

The rotator spin model is obtained from the initial phase space by
imposing the constraint
\begin{equation}
\psi = \frac{1}{2}(\bm S^2 - \rho^2) = 0, \qquad \rho > 0, \label{r4}
\end{equation}
fixing the spin of the system.
Using the Dirac method, we quantize the model in the following way.
The state space is a space of the square integrable functions on the
two--dimensional sphere. The scalar product is
$
(\Phi_1, \Phi_2) = \int_{S^2} \overline{\Phi_1(\varphi, \vartheta)}
\Phi_2(\varphi, \vartheta) \sin \vartheta d\vartheta d\varphi.
$
The above mentioned parameterization
allows us to use as the operator $\hat {\bm S}$
the usual orbital angular momentum operator expressed via spherical
angles. The wave functions as the functions on a sphere are
decomposable over the complete set of the spherical harmonics:
$
\Phi(\varphi, \vartheta) =
\sum_{l=0}^{\infty} \sum_{m=-l}^{l} \Phi_{lm} Y^l_m(\varphi, \vartheta),
$
and, therefore, the quantum analog of the first class constraint (\ref{r4}),
\begin{equation}
(\hat {\bm S}{}^2 - \rho^2) \Phi_{phys} = 0, \label{r9}
\end{equation}
leads to the quantization condition for the constant $\rho$:
\begin{equation}
\rho^2 = n(n+1), \label{r10}
\end{equation}
where $n > 0$ is an integer.
Only in this case equation (\ref{r9}) has a nontrivial solution of the form
$
\Phi_{phys}^n(\vartheta, \varphi) = \sum_{m=-n}^n \Phi_{nm} Y^n_m
(\varphi, \vartheta),
$
i.e., with the choice of (\ref{r10}) we get the states with spin equal to $n$:
$
\hat{\bm S}{}^2 \Phi_{phys}^n = n(n+1) \Phi_{phys}^n.
$
Thus, we conclude that the Dirac quantization leads to the quantization
(\ref{r10}) of the parameter $\rho$ and, as a result, the quantum system
describes the states with integer spin $n$.

Let us turn now to the construction of the reduced phase space of the system.
The constraint surface of the model can be considered as a set composed of
the points specified by two orthonormal three--vectors. Each pair of
such vectors can be supplemented by a unique third three--vector,
defined in such a way that we get an oriented orthonormal basis in three
dimensional vector space. It is well known that the set of all oriented
orthonormal bases in three dimensional space can be smoothly parameterized
by the elements of the Lie group SO(3). Thus, the
constraint surface in our case is diffeomorphic to the group manifold of
the Lie group SO(3).

The one--parameter group of canonical
transformations, generated by the constraint $\psi$,
acts in the following way:
$
\bm q(\tau) = \bm q \cos(S \tau) +
(\bm S \times \bm q) S^{-1}\sin(S \tau),
$
$
\bm S(\tau) = \bm S,
$
where
$
S =\sqrt{\bm S^2}.
$
Hence, we see that the gauge transformations are the
rotations about the direction, given by the spin vector. Thus, in the case
of a general position the orbits of the one--parameter group of
transformations under consideration are one dimensional spheres. Note,
that only the orbits, belonging to the constraint surface where $S = \rho
\ne 0$, are interesting to us. It is clear that an orbit is uniquely
specified by the direction of the spin three--vector $\bm S$ whose length
is fixed by the constraint $\psi$. As a result of our consideration, we
conclude that the reduced phase space of the rotator spin model is the
coset space SO(3)/SO(2), which is diffeomorphic to the two--dimensional
sphere $S^2$.  Due to the reasons discussed for the preceding
model there is no gauge condition in this case either. In fact, since
SO(3) is a nontrivial fiber bundle over $S^2$, we can neither find a
mapping from $S^2$ to SO(3) whose image would be diffeomorphic to the
reduced phase space. In other words, in this case the reduced phase space
cannot be considered as a submanifold of the constraint surface.

Our next goal is to write an expression for the symplectic two--form on
the reduced phase space. We can consider the variables
$S_i$ as dependent coordinates in the reduced phase space, and the
symplectic two--form on it may be expressed in terms of them.
With the help of an orthonormal basis formed by
the vectors $\bm q$, $\bm s = \bm S/S$ and $\bm q \times \bm s$,
we get for the symplectic two-form on the reduced phase space the
following expression \cite{PR}:
\begin{equation}
\omega = - \frac{1}{2\rho^2} (\bm S \times d \bm S) \wedge d \bm S.
\label{omega1}
\end{equation}
Thus, we see that the dependent coordinates $S^i$
in the reduced phase space of the system provide a realization of the basis
of the Lie algebra so(3):
\begin{equation}
\{S_i, S_j\} = \epsilon_{ijk} S_k.
\label{sss}
\end{equation}

The quantization on the reduced phase space can be performed with the help
of the geometric quantization method proceeding from the classical relations
(\ref{omega1}),
(\ref{sss}) and $\bm S^2 = \rho^2$.
This was done in detail, e.g., in
Ref.~\cite{pl6}, and we write here the final results of
this procedure. The constant $\rho$ is quantized:
\begin{equation}
\rho = j, \qquad 0 < 2j \in {\bf Z}, \label{r22}
\end{equation}
i.e., it can take only integer or half-integer value,
and the Hermitian operators, corresponding to the components of the spin
vector, are realized in the form:
$
\hat S_1 =\frac{1}{2}(1-z^{2})d/dz+jz,
$
$
\hat S_2 =\frac{i}{2}(1+z^{2})d/dz -ijz,
$
$
\hat S_{3} =zd/dz-j,
$
where
$
z = e^{-i\varphi} \tan \vartheta/2,
$
or, in terms of the dependent coordinates,
$
z = (S_1 - iS_2)/(\rho~+~S_3).
$
Operators $\hat{S}_{i}$ act in the space of holomorphic functions $f(z)$
with the scalar product
$
(f_1,f_2) = \frac{2j+1}{\pi} \int\int \overline{f_1(z)}f_2(z)(1 +
\vert z\vert^{2})^{-(2j+2)} d^{2}z,
$
in which the functions
$
\psi^{m}_{j}\propto z^{j+m},
$
$
m=-j,-j+1,...,j,
$
form the set of eigenfunctions of the operator $\hat S_{3}$ with the
eigenvalues $s_{3}=m$.
These operators satisfy the relation
$
\hat{\bm S}{}^2 = j(j+1),
$
and, therefore,
we have the $(2j+1)$--dimensional irreducible representation $D_j$ of the
Lie group SU(2).

Thus, we see that for the rotator spin model the reduced phase space
quantization method leads to the states with integer or
half--integer spin, depending on the choice of the quantized parameter
$\rho$, and gives in general the results physically different from the
results obtained with the help of the Dirac quantization method.
Let us stress once again here that within the Dirac quantization method
in this model the spin operator $\hat{\bm S}$ has a nature of
the orbital angular momentum operator,
and it is this nature that does not allow spin to
take half-integer values \cite{bied}.

\nsection{Top Spin Model}

Let us consider now the top spin model \cite{ply5}. The initial phase
space of the model is described by the spin three--vector $\bm S$, and by
three vectors $\bm e_i$ such that
$
\bm e_i \bm e_j = \delta_{ij},
$
$
\bm e_i \times \bm e_j =
\epsilon_{ijk} \bm e_k.
$
Denote the components of the vectors $\bm e_i$
by $E_{ij}$. The components
$S_i$ of the vector $\bm S$ and the quantities $E_{ij}$ form a set of
dependent coordinates in the phase space of the system. The corresponding
Poisson brackets are
\begin{equation}
\{E_{ij}, E_{kl}\} = 0,\quad
\{S_i, E_{jk}\} = \epsilon_{ikl}E_{jl}, \quad \{S_i, S_j\} =
\epsilon_{ijk} S_k.\label{5.3}
\end{equation}
The vectors $\bm e_i$ form a right orthonormal
basis in ${\bf R}^3$. The set of all such bases can be identified with the
three--dimensional rotation group. Taking into account
Eqs.~(\ref{5.3}) we conclude that the initial phase space
is actually the cotangent bundle $T^*{\rm SO(3)}$, represented as the
manifold ${\bf R}^3 \times {\rm SO(3)}$.
Using Eqs.~(\ref{5.3}), one can get the
following expression for the symplectic two--form $\omega$ on the initial
phase space:
$
\omega = \frac{1}{2} d (\bm S \times \bm e_l) \wedge d \bm e_l =
\frac{1}{2}d(\epsilon_{ijk} S_j E_{lk})\wedge d E_{li}.
$

It is useful to introduce the variables $J_i = \bm e_i \bm S = E_{ij} S_j$.
For these variables we have the following Poisson brackets:
$
\{J_i, E_{jk}\} = - \epsilon_{ijl} E_{lk},
$
$
\{J_i, J_j\} =
-\epsilon_{ijk} J_k.
$
Note, that we have the equality
$
S_i S_i = J_i J_i.
$

The phase space of the top spin model is obtained from the phase space,
described above, by introducing two first class constraints
\begin{equation}
\psi = \frac{1}{2}(\bm S^2 - \rho^2) = 0,\qquad
\chi = \bm S \bm e_3 - \kappa = 0,
\end{equation}
where
$\rho > 0,$
$|\kappa| < \rho.$
Consider now the Dirac quantization of the model.
Let us parameterize the matrix $E$, which can be identified
with the corresponding rotation matrix,
by the Euler angles, $E = E(\alpha,
\beta, \gamma)$, and use the representation where the operators,
corresponding to these angles are diagonal. In this representation state
vectors are functions of the Euler angles, and the operators $\hat S_i$ and
$\hat J_i$ are realized as linear differential operators, acting on such
functions \cite{var}. The quantum analogs of the constraints $\psi$ and
$\chi$ turn into the equations for the physical states of the system:
\begin{equation}
(\hat{\bm S}{}^2 - \rho^2) \Phi_{phys} = 0,\qquad
(\hat J_3 - \kappa) \Phi_{phys} = 0.\label{5.10}
\end{equation}
An arbitrary state vector can be decomposed over the set of the Wigner
functions, corresponding to either integer or half--integer spins
\cite{var}:
$
\Phi(\alpha, \beta, \gamma) = \phi_{jmk} D^j_{mk}(\alpha, \beta, \gamma),
$
where $j = 0, 1, \ldots$, or $j = 1/2, 3/2, \ldots$, and $k, m = -j, -j+1,
\ldots, j$. The Wigner functions $D^j_{mk}$ have the properties:
$
\hat{\bm S}{}^2 D^j_{mk} = j(j+1) D^j_{mk},
$
$
\hat S_3 D^j_{mk} = m D^j_{mk},
$
$
\hat J_3 D^j_{mk} = k D^j_{mk}.
$
Using the decomposition of the state vector, we see that
Eqs.~(\ref{5.10}) have nontrivial solutions only when
$\rho^2 = j(j+1)$, and $\kappa = k$, for some integer or half--integer
numbers $j$ and $k$, such that $-j \le k \le j$. In other words we get the
following quantization condition for the parameters of the model:
\[
\rho^{2} = j(j+1),\quad
\kappa = k,\qquad
-j \le k \le j,\quad
0 < 2j \in{\bf Z}.
\]
The corresponding physical state vectors have the form
\[
\Phi_{phys}(\alpha,\beta,\gamma) = \sum_{m=-j}^j \varphi_m
D^j_{mk}(\alpha,\beta,\gamma).
\]
Thus, we see that the Dirac quantization of the top spin model leads to
an integer or half--integer spin system.

Proceed now to the construction of the reduced phase space of the system.
As the constraints $\psi$ and $\chi$ have zero Poisson bracket, we can
consider them consecutively. Let us start with the constraint $\psi$.
{}From the expressions for the Poisson brackets (\ref{5.3})
it follows that the group of gauge transformations, generated by the
constraint $\psi$, acts in the initial phase space variables as follows:
\[
\bm e_i(\tau) = \bm e_i \cos (S \tau) + (\bm S \times \bm e_i) S^{-1}
\sin (S \tau) + \bm S (\bm S \bm e_i)S^{-2} (1 - \cos (S \tau)),
\quad
\bm S (\tau) = \bm S,
\]
where $S = \sqrt{\bm S^2}$.
We see that the transformation under consideration have the sense of the
rotation by the angle $S\tau$ about the direction of the spin vector.
Let us consider
the initial phase space of the system being diffeomorphic to
${\bf R}^3 \times {\rm SO(3)}$
as a trivial
fibre bundle over ${\bf R}^3$ with the fibre SO(3).
The gauge transformations
act in fibres of this bundle. It is
clear that the constraint surface, defined by the constraint $\psi$, is a
trivial fibre subbundle $S^2 \times {\rm SO(3)}$. As ${\rm SO(3)/SO(2)} =
S^2$, then after the reduction over the action of the gauge group we come
to the fibre bundle over $S^2$ with the fibre $S^2$. As it follows from
general theory of fibre bundles \cite{Hus66}, this fibre bundle is again
trivial. Thus the reduced phase space, obtained using only the constraint
$\psi$, is the direct product $S^2 \times S^2$. The
symplectic two--form on this reduced space can be written in the form
\cite{PR}:
$
\omega = - (2 \rho^{2})^{-1} (\epsilon_{ijk} S_i dS_j \wedge dS_k -
\epsilon_{ijk} J_i dJ_j \wedge dJ_k).
$
Here the quantities $S_i$ and $J_i$ form a set of
dependent coordinates in the reduced phase space under consideration:
$S_i S_i = J_i J_i = \rho^2$.

Let us turn our attention to the constraint $\chi$. It is easy to get
convinced that the transformations of the gauge group, generated by this
constraint act in the initial phase space in the following way:
$
\bm e_i(\tau)=\bm e_i \cos\tau+(\bm e_3\times \bm e_i)\sin\tau,
$
$i=1,2,$
$
\bm e_3(\tau)=\bm e_3,
$
$
\bm S(\tau) = \bm S.
$
So, we see that the gauge group, generated by the constraint $\chi$, acts only
in one factor of the product $S^2 \times S^2$, which is a reduced phase
space obtained by us after reduction with the help of the constraint
$\psi$.  Thus we can consider only that factor, which is evidently
described by the quantities $J_i$. From such point of view, the constraint
surface, defined by the constraint $\chi$, is a one dimensional sphere
$S^1$, where the group of gauge transformations acts transitively.  Hence,
after reduction we get only one point. Thus, the final reduced phase space
is a two--dimensional sphere $S^2$, and
the symplectic two--form on the reduced phase space has the form given by
Eq. (\ref{omega1}).
Therefore, the reduced phase space we have obtained,
coincides with the reduced phase
space for the rotator spin model. Hence the geometric quantization method
gives again the quantization condition  (\ref{r22}) for the parameter $\rho$,
while the parameter $\kappa$ remains unquantized here. Therefore, while
for this model unlike the previous one, two methods of quantization lead
to the quantum system, describing either integer or half-integer spin
states, nevertheless, the corresponding quantum systems are different: the
Dirac method gives discrete values for the observable $\hat J_3$, whereas
the reduced phase space quantization allows it to take any value $\kappa$,
such that $\kappa^2 < j^{2}$ for a system with spin $j$.

Let us note here one interesting property of the system.  We can use a
combination of the Dirac and reduced phase space quantization methods.
After the first reduction  with the help of the constraint $\psi$, the system,
described  by the spin vector and the `isospin' vector \cite{ply5} with
the components $I_i = -J_i$, $S_i S_i = I_i I_i$, can be quantized
according to Dirac by imposing the quantum analog of the constraint $\chi$
on the state vectors for singling out the physical states.  In this case
we have again the quantization of the parameter $\kappa$ as in the pure
Dirac quantization method, and, therefore, here the observable $\hat J_3$
can take only integer or half--integer value.  Hence, in this sense, such
a combined method gives the results coinciding with the results of the
Dirac quantization method.

\nsection{Discussion and conclusions}

The first considered model
gives an example of the classical
constrained system with finite number of the degrees of freedom for which
there is no gauge condition, but nevertheless, the reduced phase space can
be represented as a submanifold of the constraint surface.
As we have seen, Dirac and reduced phase space quantization methods lead to
the coinciding physical results for this plane spin model.
Moreover, we have revealed
an interesting analogy in interpretation of the
situation with nonexistence of a global gauge condition for this simple
constrained system with the situation taking place for the non-Abelian
gauge theories \cite{sin}.

The rotator and top spin models
give examples of the classical systems, in which
there is no global section of the space of gauge orbits. In
spite of impossibility to impose gauge conditions such systems admit the
construction of the reduced phase space.
These two models demonstrate that the reduced
phase space and the Dirac quantization methods
can give essentially different physical results.

Thus, for
Hamiltonian systems with first class constraints we encounter two related
problems.

The first problem consists in the choice of a `correct'
quantization method for such systems. From the mathematical point of view
any quantization leading to a quantum system, which has the initial system
as its classical limit, should be considered as a correct one, but
physical reasonings may distinguish different quantization methods.
Consider, for example, the above mentioned systems. The rotator spin
model, quantized according to the Dirac method, represents by itself the
orbital angular momentum system with additional condition (\ref{r9})
singling out the states with a definite eigenvalue of angular momentum
operator $\hat{\bm S}{}^{2}$. This eigenvalue, in turn, is defined by the
concrete value of the quantized parameter of the model: $\rho^2 =
n(n+1)>0$.  On the other hand, the reduced phase space quantization of the
model gives either integer or half--integer values for the spin of the
system. If we suppose that the system under consideration is to describe
orbital angular momentum, we must take only integer values for the
parameter $\rho$ in the reduced phase space quantization method.
But in this case we must, nevertheless, conclude, that the reduced phase
space quantization method of the rotator spin model describes a more general
system than the quantum system obtained as a result of the Dirac
quantization of that classical system.

The Dirac quantization of the top spin model, or its combination
with the reduced phase space quantization gives us a possibility to
interpret this system as a system having spin and isospin degrees of
freedom (with equal spin and isospin: $\hat{\bm S}^2=\hat{I}_i\hat{I}_i =
j(j+1)$), but in which the isospin degrees of freedom are
`frozen' by means of the condition $\hat{I}_{3}\Phi_{phys} = -k
\Phi_{phys}$.  On the other hand, as we have seen, the reduced space
quantization method does not allow one to have such interpretation of the
system since it allows the variable $I_{3}$ to take any (continuous) value
$-\kappa$ restricted only by the condition $\kappa^2 < j^2$, i.e.,
the operator $\hat{I}_{3}$ (taking here only one value) cannot be
interpreted as a component of the isospin vector operator.
{}From this point of view a `more correct' method of quantization is the
Dirac quantization method.

In this respect it is worth to point out that there is
a class of physical models, for which it is impossible to get
the reduced phase space description, and which, therefore, can be
quantized only by the Dirac method.

Indeed, there are various pseudoclassical models containing first class
nilpotent constraints of the form \cite{spin1}--\cite{cor}:
\begin{equation}
\psi = \xi_{i_{1}}...\xi_{i_{n}}G^{i_{1}...i_{n}} = 0, \label{7.1}
\end{equation}
where $\xi_{i_{k}}$, are real Grassmann variables with the Poisson brackets
$
\{\xi_{k},\xi_{l}\}=-ig_{kl},
$
$g_{kl}$ being a real nondegenerate symmetric constant matrix. Here
it is supposed that
$G^{i_{1}...i_{n}}$, $n \ge 2$, are some functions of other variables,
antisymmetric in their indices, and
all the terms in a sum have simultaneously either even or odd
Grassmann parity.
For our considerations it is important that constraints
(\ref{7.1}) are the constraints, nonlinear in Grassmann variables,
and that they have zero projection on the unit of Grassmann algebra.
In the simplest example of relativistic massless vector particle in
(3+1)--dimensional space--time \cite{spin1} the odd part of the
phase space is described by two Grassmann vectors $\xi_{\mu}^{a}$,
$a=1,2$, with brackets
$
\{\xi_{\mu}^{a},\xi_{\nu}^{b}\}=-i\delta^{ab}g_{\mu\nu},
$
and the corresponding nilpotent first class constraint has the form:
\begin{equation}
\psi = i\xi_{\mu}^{1}\xi_{\nu}^{2}g^{\mu\nu} = 0, \label{7.4}
\end{equation}
where $g_{\mu\nu}= \mbox{diag}(-1,1,1,1)$.
This constraint is the generator of
the SO(2)--rotations in the `internal isospin' space:
$
\xi_\mu^1 (\tau) = \xi_\mu^1 \cos \tau + \xi_\mu^2 \sin \tau,
$
$
\xi_\mu^2 (\tau) = \xi_\mu^2 \cos\tau - \xi_\mu^1 \sin \tau.
$
The specific property of this transformation is
that having $\xi^a_\mu(\tau)$ and $\xi_\mu^a$, we cannot determine the
rotation angle $\tau$ because there is no notion of the inverse element
for an odd Grassmann variable.  Another specific feature of the nilpotent
constraint (\ref{7.4}) is the impossibility to introduce any, even local,
gauge constraint for it. In fact, we cannot find a gauge constraint $\chi$
such that the Poisson bracket $\{\psi,\chi\}$ would be invertible.
Actually, it is impossible {\it in principle} to construct the
corresponding reduced phase space for such a system.
Obviously, the same situation arises for the constraint of general form
(\ref{7.1}).  It is necessary to note here that in the case when the
constraint $\psi$ depends on even variables of the total phase space (see,
e.g., ref. \cite{cor}), and, therefore, generates also transformations of
some of them, we cannot fix the transformation parameter (choose a
point in the orbit) from the transformation law of those even variables,
because the corresponding parameter is present in them with a
noninvertible factor, nonlinear in Grassmann variables.  Therefore, the
pseudoclassical systems containing the constraints of form (\ref{7.1}) can
be quantized only by the Dirac method, that was done in
original papers \cite{spin1}--\cite{cor}.

Let us come back to the discussion of the revealed difference
between two methods of quantization, and point out that
the second related problem is clearing up the sense of gauge degrees of
freedom. The difference appearing under the Dirac and reduced phase
space quantization methods can be understood as the one proceeding from the
quantum `vacuum' fluctuations corresponding to the `frozen' (gauge)
degrees of freedom.  Though these degrees of freedom are `frozen' by the
first class constraints, they reveal themselves through quantum
fluctuations, and in the Dirac quantization method they cannot be
completely `turned off' due to the quantum uncertainty principle. Thus, we
can suppose that the gauge degrees of freedom serve not simply for
`covariant' description of the system but have `hidden' physical meaning,
in some sense similar to the compactified degrees of freedom in the
Kaluza--Klein theories. If we adopt such a point of view, we have to use
only the Dirac quantization method.  Further, the gauge principle cannot
be considered then as a pure technical principle. From here we arrive also
at the conclusion that the Dirac separation of the constraints into first
and second class constraints is not technical, and nature `distinguish'
these two cases as essentially different, since gauge degrees of freedom,
corresponding to the first class constraints, may reveal themselves at the
quantum level (compare with the point of view advocated in Ref.~\cite{jac}).

$\ $

The work of M.P. was supported in part by MEC-DGICYT, Spain.

\end{document}